# Design and Application of Data Aquistion Interface Circuit

**Hayder O. Alwan\* Noor M. Farhan, Qais S- Al-Sabbagh**

*University of Baghdad*

*Hayderalwan1981@gmail.com:*

## Abstract

A commitment to condition monitoring involves the operators of plant in the conduct of a range of activities. These activities may be complicated in nature and indeed may often be performed automatically under computer control. They can, however, always be down into a relatively small number of easily identifiable functional tasks. This makes it much easier to identify the common elements of machine condition monitoring schemes. A proposed interface circuit design and application will be further explain in this paper, the implemented monitoring unit circuit also illustrated, see appendix A. Two scenarios presented in this paper, first ten turns assume to be shorted, and in the second thirty turns shorted to show the difference in the amplitude of frequencies at each case. This paper present. An improvement in three-phase squirrel-cage induction motor stator inter-turn fault detection and diagnosis based on a neural network approach is presented.

## 1. Introduction

Outages and faults cause problems in interconnected power system with huge economic consequences in modern societies [1]. Especially with moving forward to modern electricity network, a significant study has been done for address different operation status of power systems. For instance, With the increasing concerns over environmental emissions from fossil fuel based electric power generation, the utilization of energy efficient CHP (combined heat and power) systems to simultaneously meet electricity and thermal demands can be economically rewarding, while assisting with reducing environmental emissions. To address that, reference [2] developed and simulated a novel approach for optimal economic dispatch scheduling for a GENCO to maximize economic profit and minimize environmental emissions based on integration of CHP systems with conventional thermal power generating units, where a double Benders decomposition solution approach is proposed for optimization. Another important aspect of power systems are Electric machines. Electric machines could play a significant role in restoration procedure. Therefore their simulations on different fault situations is important. The aim of this Paper is to:

1- Motor current Signature Analysis applied in order to monitor and detect the Inter-turn fault Motor.

2- This paper presents the design and realization of data acquisition, as monitoring system for the interference between the PC lab and the motor.

3- Training the neural network to simulate the effects of three types of 3-phase induction machine faults. The required training data which will be used to train the ANN are obtained practically by the designed monitoring system [3].

II. SHORT CIRCUIT IN STATOR WINDING

In the long term, the multiple stresses cause ageing, which finally leads to insulation breakdown. For this reason, it is important to estimate the remaining insulation integrity of the winding after a period of operating time [4].

.This will be of little consequence but it will be quantifiable in the flux distribution in the air-gab. [5]

The frequency components to be detected in the axial flux component is given by,

$$(k \; n(1-s)/p) \; f \tag{1}$$

where p is the number of pole pairs, $f$ is the mains frequency, k=1,3 and n = 1,2,3,….,(2p-1) and s is the slip.

Fig.1 shows an inter-turn short circuit between two points,a and b,of a complete stator winding .

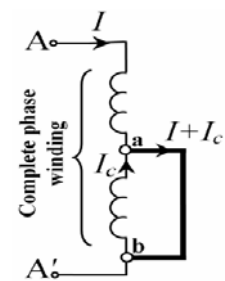

Fig. (1) Inner-turn short circuit

## II. FUNDAMENTALS OF MONITORING SYSTEMSUSING (MCSA

A commitment to condition monitoring involves the operators of plant in the conduct of a range of activities. These activities may be complicated in nature and indeed may often be performed automatically under computer control. They can, however, always be down into a relatively small number of easily identifiable functional tasks. This makes it much easier to identify the common elements of machine condition monitoring schemes. The present paper discusses the fundamentals of Motor Current Signature Analysis (MCSA) plus condition monitoring of the induction motor using MCSA [4].

## III. ELEMENTS OF A MONITORING SYSTEM

A sophisticated monitoring system can read the entrances of hundreds of sensors and execute mathematical operations and process a diagnosis. Currently, the diagnosis is gotten, most of the time, using artificial intelligence techniques, a monitoring system can be divided in four main stages [6]:

• Transduction of the interest signals;

• Acquisition of the data;

• Processing of the acquired data;

• Diagnosis.

Figure (2) presents a pictorial form of this process.

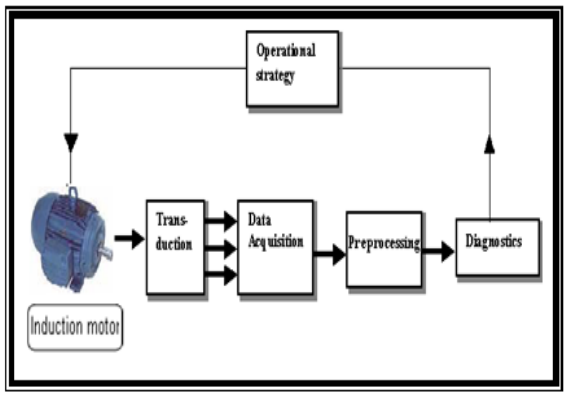

Fig. (2) Condition Monitoring System

The result of this is that, with respect to the stationary stator winding, this backward rotating field at slip frequency with respect to the rotor induces voltage and current in the stator winding at [7]

fsb=f1(1-2s)Hz                                              (1)

f1=supply frequency, Hz, f2= sf1 Hz, f2= slip frequency, Broken rotor bars therefore result in current components being induced in the stator winding at frequencies given by:

fsb=f1(1±2s)Hz                                            (2)

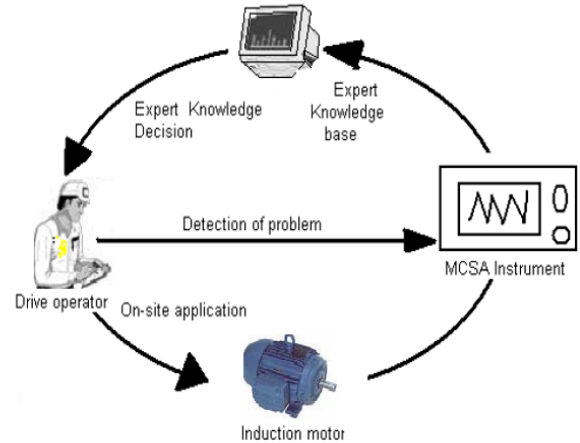

**Fig. (3)** Overall MCSA Strategy

## IV. EXPERIMENTAL WORK AND RESULTS

Fig .( 4) and Fig. (5) illustrate  the testbed used in this work. The system consists of  2.2KW Motor , 50 Hz, 2 pole, 3000 rpm induction motors, pulleys, belt, computer with Data Acquisition, Oscilloscope and spectrum analyzer used to create the data needed under no-load conditions. A dc Generator of 3KW is coupled with the motor by the pulley and belt as shown in Fig.(5). The fundamental operation of the Data Acquisition system, including detailed descriptions of the measurment system and control systems will be disccussed in this paper.

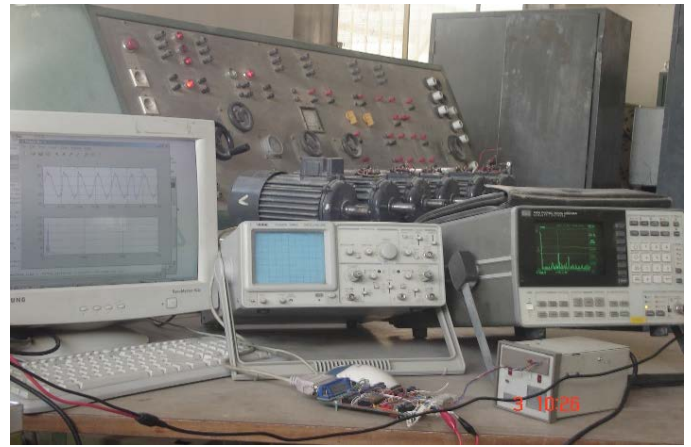

**Fig.(4)** Close Up Of Laboratory Equipments Set-Up

Inter -turn performed by two different numbers of short-circuited stator turns to achieve two different degrees of faults as can be seen in Fig. (6).Three taps were made in stator turns (10 turns) between the first and the second tap. (20 turns) between the second tap and the third tap. There are 67turns in

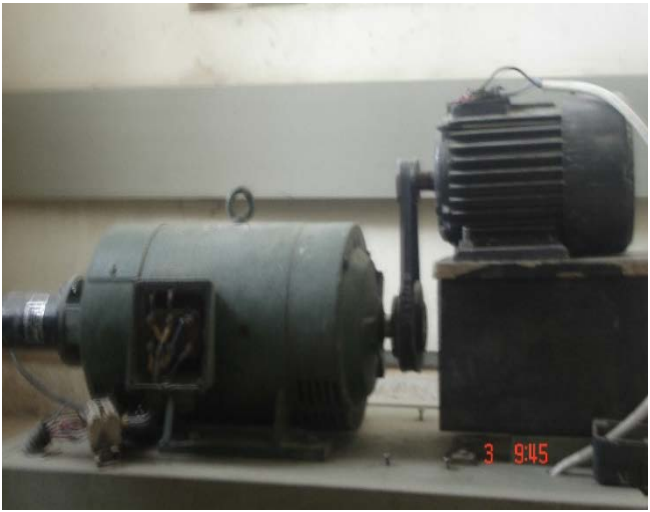

Fig.(5) Flexible Coupling

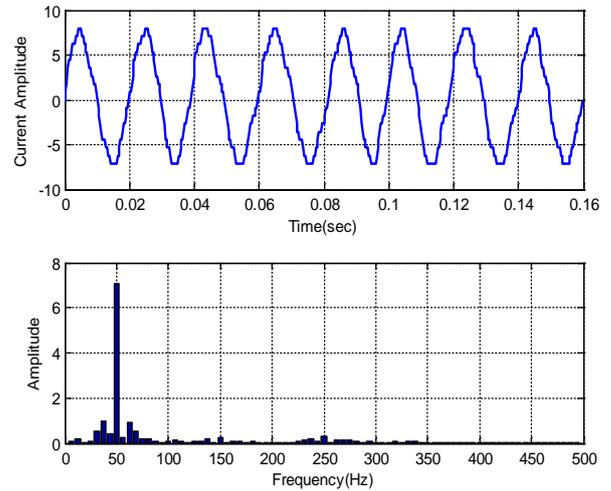

**Fig. (7)** Current waveform of stator inter-turn fault at no-load
a)Line  current wavefom                b) FFT

Each coils. The FFT analyses were performed on the acquired data. The side bands frequencies and their amplitudes was calculated as shown in tables by applying equation (2.3) in two cases both these cases at no load.

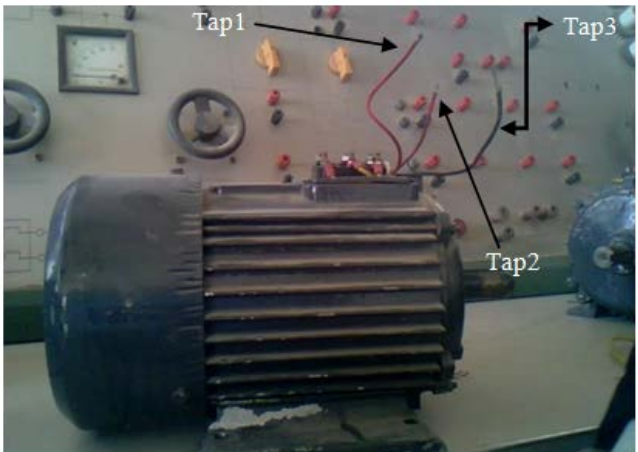

Fig.(6) Modeled stator inter-turn fault

**VI.A. (10 Turns Shorted At No-Load Test)** In this case the shorten was between the tap1 and tap 2, the number of shorted turns was ten. The stator line current and it's FFT as shown in Fig. (7) The current and the speed was 8 A.2650rpm respectively.

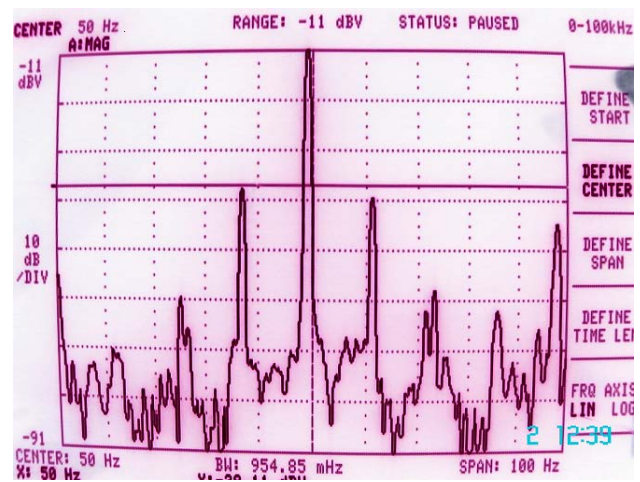

Table 4.10 Illustrates the positive negative harmonics sequence and their amplitudes for different values of $k$ at no-load,the data of the motor
is:
 Case 1.when the shorted turn from (10-20).
 S=ns-nr/ns= 0.133 ,nr=2650
$f_s=(k \pm n(1-s)/p)f$
Where  p is the number of pole pairs, $f$ is  the mains frequency , n=1

### VI.B. (30 Turns Shorted At No-Load Test )

In this case the shorte was between the tap1and tap 3 that means the number of shorted turns was thirty.The stator line current and it's FFT as shown in Fig.(4-34). The current speed was 9 A.2500 rpm

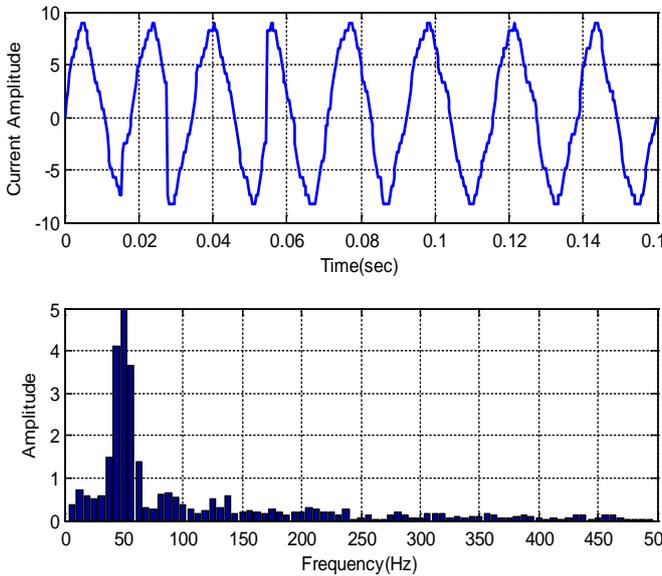

**Fig. (9)** Current waveform of stator inter-turn fault at no-load
a)Line current wavefom                    b) FFT

**Table 4.11** Illustrates the positive negative harmonics sequence and their amplitudes for different values of *k* at no load,the data of the motor is:

| k | Pos.Harmonic | Amplitude | Neg.Harmonic | Amplitude |
|---|---|---|---|---|
| 1 | 92 | 0.1312 | 8 | 0.396 |
| 3 | 194 | 0.071 | 67 | 0.395 |
| 5 | 292 | 0.037 | 125 | 0.23 |
| 7 | 392 | 0.011 | 183 | 0.05 |
| 9 | 492 | 0.029 | 242 | 0.062 |
| 11 | 592 | 0.005 | 300 | 0.05 |
| 13 | 692 | 0.013 | 385 | 0.041 |
| 15 | 792 | 0.014 | 416 | 0.004 |
| 17 | 892 | 0.0184 | 475 | 0.053 |
| 19 | 992 | 0.01 | 533 | 0.027 |
| 21 | 1092 | 0.03 | 591 | 0.028 |

Case 2.when the shorted turn from (10 -40).
  S=ns-nr/ns=0.166,nr=2500

$f_s=(k\pm n(1-s)/p)f$

Where p is the number of pole pairs, *f* is the mains frequency ,n=1

| Inupt Frequency | Motor Speed | Slip (s) | KW Rating | pole pairs |
|---|---|---|---|---|
| 50 H | 2500rpm | 0.166 | 2.2kw | 1 |

| k | Neg.Harmonic | Amplitude | Pos.Harmonic | Amplitude |
|---|---|---|---|---|
| 1 | 6 | 0.323 | 94 | 0.01288 |
| 3 | 106 | 0.1119 | 194 | 0.0126 |
| 5 | 206 | 0.0176 | 294 | 0.018 |
| 7 | 306 | 0.0078 | 394 | 0.0156 |
| 9 | 406 | 0.0126 | 494 | 0.0139 |
| 11 | 506 | 0.0151 | 594 | 0.0056 |
| 13 | 606 | 0.0103 | 694 | 0.0119 |
| 15 | 706 | 0.0033 | 794 | 0.0082 |
| 17 | 806 | 0.0147 | 894 | 0.01 |
| 19 | 906 | 0.015 | 994 | 0.0013 |

### VII. THE PROPOSED CIRCUIT OPERATION

The four test signals are applied to the four inputs (in1- in4) of the IC7, which is dual (4-1) line analog switch multiplexer IC, the four inputs of this IC are at pins (1,5,2,4). The selection of the input lines is controlled by the control pins (10,9) which are (A,B) control lines, these lines are controlled by the LPT bus of the microcomputer. The signal are required to be made in a suitable form before providing them to analog multiplexer ; this can be done using the following simple circuits:

1. Speed signal Voltage circuit, this circuit reduce the voltage (speed) by factor (1/18 ) uses the resistors (R13 & R14), shown in Fig. (4-4) , this circuit used the range of analog voltage signal coming from Tachometer from range (0-90V) to range (0-5V).
2. Motor current circuit, by using the circuit shown in Fig. (4-4) to invert of signal polarity .This circuit used to invert the current signal coming from CT to the voltage signal before providing it to the multiplexer circuit.
3. Voltage follower (buffer) circuit, the circuit shown in Fig. (4-4) is used because its input resistance is high . Therefore, it draws negligible current from signal source, and then the loading effect will be removed. This circuit is used in different place of the system hardware, such as using it between the output of the multiplexer circuit and the input of analog to digital converter. Pin (3) of the IC7 is the output pin that accessed to the input of the sample and hold IC1, this IC converts the analog signal to the sampled signal , where, the hold time depends on capacitance value of the capacitor C1.The sampling signal comes from the invertion of the STS output pin of the Analog-to-digital convertor IC2, its frequency value is choosed by the software user's. The output of the sampler IC1 is accessed to the input of the Analog-to-Digital convertor IC2. This IC has two convertion modes these are: (8bits, 12bits) conversions, the first one was selected.

In this mode the control pins (12/8, A$_o$, CE) or (2,4,6) connected to the +Vcc, and (Analog com, CE ) or (pins 9, 3) connected to the ground (GND) pin. The output data bits of IC2 are (D4-D11) at pins(20-27).The AD574 type was chosen for IC2 in this design, where this IC converts the sampled analog signal to digital numbers, the lowest level of the analog is converted to (00)$_H$ number ,while the highest level is converted to (FF)$_H$ number. The output data bits (D4-D11) are connected to the 8-bits input lines of the Latch circuit(IC3 type 74HC374), this IC saves the output digital number of the A/D convertor for a period of time while the microcomputer read this data from the LPT bus (pins 10,11,12,13).The loading (saving) signal came from the invertion of the STS output control pin of the A/D convertor IC2. The output data from the Latch ciruit (IC3) goes to to the buffer drive circuit(IC4) type 74LS241, which it acts as a nibble (4-Bits) selector. The control pins (OEA, OEB) are shorted and connected to the pin(5) [D3] of the LPT bus, if this line goes to High state then the higher state nibble (D4-D7) is selected, and if this line goes to Low state then the Lower nibble (D0-D3) is selected. The output pins of this IC are shorted by this manner, Q0 to Q4, Q1 to Q5, Q2 to Q6, Q3 to Q7.

The output pins (Q0-Q2) or (Q4-Q6) of IC4 are connected to iput pins (Q4-Q6) of the LPT bus, while Q3 of IC4 is inverted then connected to Q7 of LPT bus. The combinational circuit of ¼ IC6, 1/6 IC5 and C2 is used as a monostable circuit (one-shot to), it converts the clock signal that came from pin(4) (D2) of the LPT bus to series of narrow Low pulses that goes to the (R/C) control pin of A/D convertor IC. The resistors (R3-R10) are used for isolation and protection of the LPT bus of the microcomputer from the proposed circuit.See Appendix B

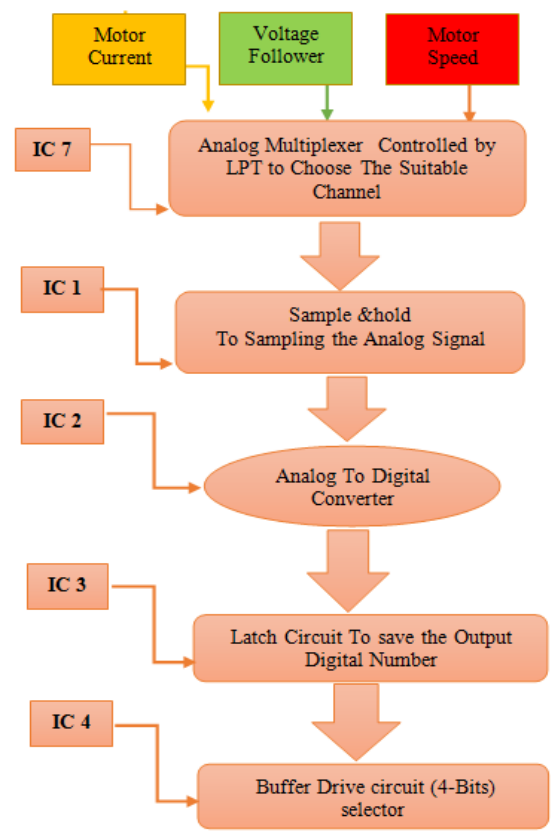

**Fig.(10)** Block diagrame of the main sections of the data acquisition system

## VIII. SOFTWARE IMPLEMENTATION

The implementation of the software programe included five steps in the first stage the data of the current and the speed given to the data acquisition circuit, the line current measured by using current transformer (10/4) A passing through a resistance of 1Ω which given 4 volt to the data acquisition circuit, this circuit will convert these analogue current signal to 3250 digtal numbers, in the stage two these numbers will loaded to the FFT programe to abtain the sampling frequency and sampling time of the waveform (see appendix c-4).

The speed of the motor measured by using the tachometer the value of the speed will convert to the voltage value, it's found that the tachometer used in the laboratory give 0.06 volt for each rotation, for example if the motor speed is 2800 rpm the tachometer give 168 volt this value will reduce to about 4 volt before used in data acqusition circuit. Stage three include the calculation of the frequencies of the positive, negative side bands by applying the equeations which are realated to the five faults, this stage also included extraction of the amplitudes of these frequencies see appendix c-2. Stage four refer to the rule of the neural network in faults detection , the amplitudes of the side bands used as inputs as mentioned before in this chapter , see the training of the neural programe in appendix c-3. Fig.(11) illustrates the flowchart of the system operations. The basic following steps illustrates the Fast Fourier Transform (FFT) which are:

:

1-  all line currents of the healthy and faulty motors should be given to the programe (see appendix c-4).
2-  Choose the number of the cycle (six cycles have been chosen in this work) as shown from line current for all type of faults .
3-  Calculate the length of the six waves.
4-  Calculate the sampling time and sampling frequency. The four steps mentioned above done using the programe in (appendix c-4).
5-  Applying the FFT package (yy1=fft(b,s)) see appendix c-5, where b is the values of the numbers by which the six cycles were drawn, s is the total number of these values it was equal to 390 for the six cyles (65 values for each cycle over 0.02 (sec)). Therefore the sampling time can be calculated by divided the 65 values on 0.02(sec) and the result is the sampling time it was equal to 3.076923e-4sec, then the sampling frequency is: 1/sampling time =3250 Hz

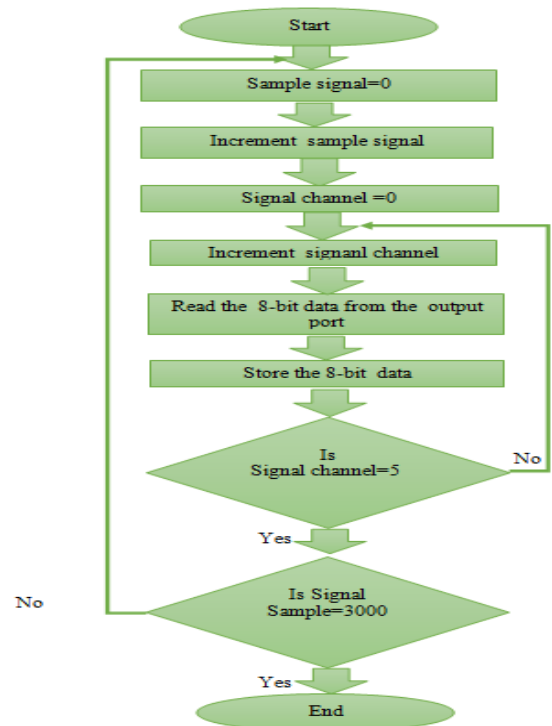

Fig. (12) Flow chart of interface program

## IX. CONCLUSION AND FUTURE WORK

The work reported in this paper has involved designing and building a motor monitoring system using an Aritificial neural network for fault detction of three phase induction motor .To accomplish this, a hardware system was designed and built to acquire three-phase stator current and speed from a (2.2kw) squirrel-cage induction motor. The ability of the phase current to detect specific fault was tested, since monitoring this parameter is the most convenient and cheapest way to sense a fault. it was clear that The sideband frequencies are function of the slip, so they are changing with the speed (that change with the load) .

From the sideband frequencies calculated in the tables(1 and 2) it's found that the distance of the positive and negative from the fundamental increased with increasing of the load, and the same for different values of k/p and for all types of faults. From the reported work,the disadvantage of most ANN's are their inability to respond to previously unseen conditions. Therefore , if there is an occurance of a new fault that the network doesn't been train to recognize ,and the fault may be misdiagnosed which produce weak output results. The number of selected input channels of data acquisition circuit used in this work was two (one for current and the other for tachometer ), these slected chanals may increase to four depend on the types of data needed to investigated.

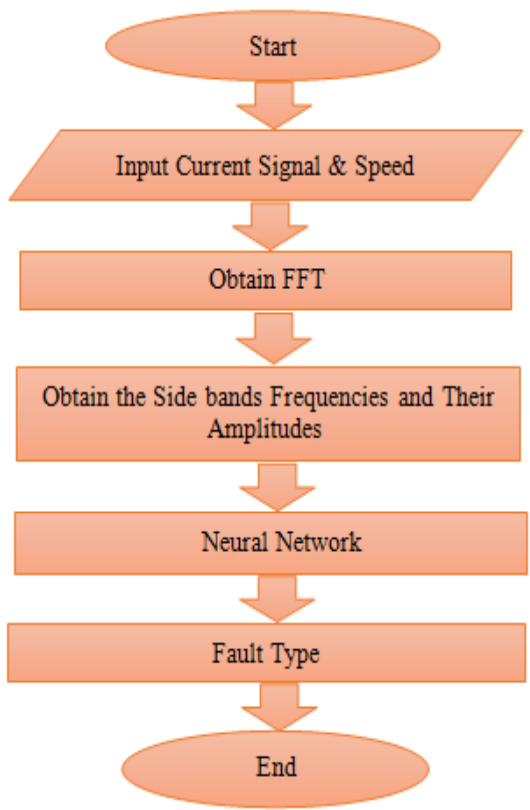

**Fig. (11)** Flowchart of system operations

Fig.(12) illustrates the flowchart of the quick basic programe used in the interface circuit, the acquisition circuit channel have been controlled by using the programe of the interfac (see appendix d).

## Acknowledgement

This work was possible due to the support given by university of Technology. Special hanks to Dr. Fadhil Abbas. Mrs. Fatimah. Mr. Khleej and Dr. Nawal .

## REFERENCES

[1]. Alwan, Hayder O., and Noor M. Farhan. "Load Restoration Methodology Considering Renewable Energies and Combined Heat and Power Systems." arXiv preprint arXiv:1806.01789 (2018).

[2]. Sadeghian, H. R., and M. M. Ardehali. "A novel approach for optimal economic dispatch scheduling of integrated combined heat and power systems for maximum economic profit and minimum environmental emissions based on Benders decomposition." Energy 102 (2016): 10-23.

[3]. S.Nandi, H.A.Toliyat, "Novel Frequency Domain_ Based Technique to Detect Incipient Stator Intertern Faults in Induction Machines", The 2002 IEEE Industry Application Society Conference, The 35th IAS Annual Meeting, Rome, ItalyVol, 38.No.1, January/February 2002.

[4]. A.Yazidi,D.Thailly,H.Henao,R.Romary,G.A.Capolino,J.F.Brudny,2004"Detection of Stator Short-Circuit in Induction Machines Using an External Leakage Flux Senser"IEEE International Conference on Industrial Technology- ICIT'Dec.8-10,2004, Hammamet, Tunisia, Vol.1,pp. 166-169.

[5]. Alwan, Hayder E., and Qais S. Al-Sabbagh. "Detection of Static Air-Gap Eccentricity in Three Phase Induction Motor by Using Artificial Neural Network (ANN)." Journal of Engineering 15.4 (2009): 4176-4192.

[6]. P. Zhang, Y. Du, G. Thomas Habetler, and Bin Lu. "A survey of condition monitoring and protection methods for medium-voltage induction motors", IEEE Trans. Ind. Appl., Vol. 47, No. 1, pp. 34-46, 2011.

[7]. Hayder O. Alwan, Noor M. Farhan, Qais S-Al-Sabbagh "Detection of Static Air-Gap Eccentricity in Three Phase induction Motor by Using Artificial Neural Network (ANN)" Vol 7-Issue 5(May-2017) *International Journal of Engineering Research and Application(IJERA)*, ISS: 2248-9622, www.ijera.com. Available online: http://www.ijera.com/papers/Vol7_issue5/Part-3/D0705031523.pdf

## Appendix A: Structure of Data Acquisition Circuit

t.

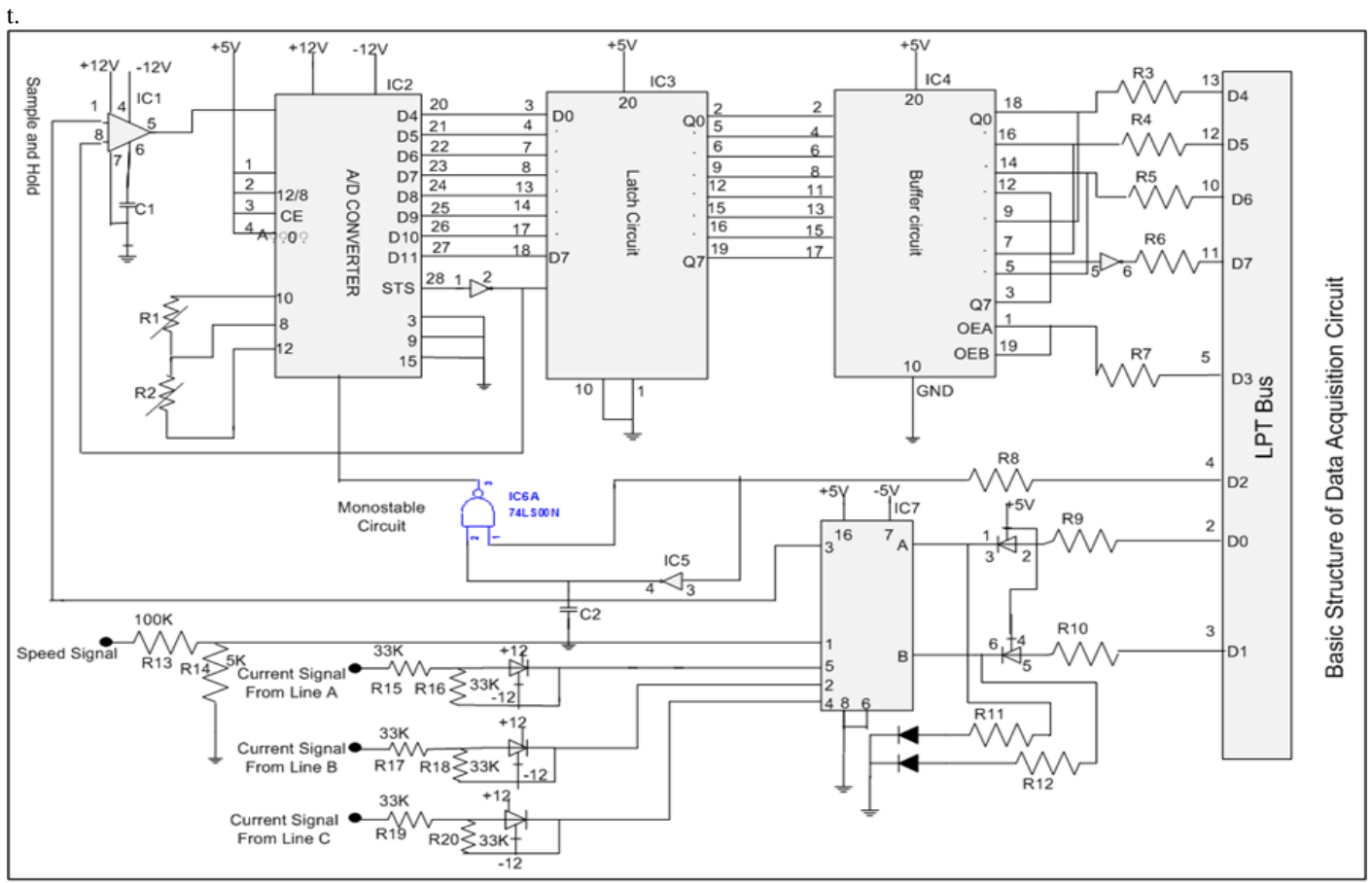